\documentclass[11pt]{article}

\usepackage[preprint]{acl}
\usepackage{times}
\usepackage{latexsym}
\usepackage{tabularx}
\usepackage[T1]{fontenc}

\usepackage[utf8]{inputenc}

\usepackage{microtype}

\usepackage{inconsolata}

\usepackage{graphicx}

\usepackage{booktabs}
\usepackage{amsmath}
\title{AudioRole: An Audio Dataset for Character Role-Playing in Large Language Models}

\author{Wenyu Li\footnotemark[1]\\
  Westlake University \\
  \And
  Xiaoqi Jiao \\
  LIGHTSPEED \\
  \AND
  Yi Chang \\
  LIGHTSPEED \\\And
  Guangyan Zhang \\
  LIGHTSPEED \\\And
  Yiwen Guo \\
  Independent Researcher \\}

\begin{document}
\maketitle
\renewcommand{\thefootnote}{\fnsymbol{footnote}}
\footnotetext[1]{This work was done when Wenyu Li (\texttt{liwenyu@westlake.edu.cn}) was an intern at Tencent LIGHTSPEED STUDIOS.}
\renewcommand{\thefootnote}{\arabic{footnote}}
\begin{abstract}
While existing role-playing research predominantly focuses on text, \textbf{Audio Role-Playing (ARP)} presents unique challenges regarding the synchronized alignment of semantic content and vocal characteristics. To address this gap, we propose \textbf{AudioRole}, a meticulously curated dataset from 13 TV series spanning 1K+ hours with 1M+ character-grounded dialogues, providing synchronized audio-text pairs annotated with speaker identities and contextual metadata. In addition, to demonstrate the effectiveness of the dataset, we introduced \textbf{ARP-Eval}, a dual-aspect evaluation framework that assesses both \textit{response quality} and \textit{role fidelity}. Empirical validation showing GLM-4-Voice trained on AudioRole (called \textbf{ARP-Model}) achieves an average Acoustic Personalization score of 0.31, significantly outperforming the original GLM-4-voice and the more powerful model MiniCPM-O-2.6. The \textbf{ARP-Model} also achieves a Content Personalization score of 0.36, surpassing the untrained original model by about 38\%. The blind human perceptual evaluation also confirms these findings.

\textbf{AudioRole} features dialogues from over 115 main characters, 6 trained \textbf{ARP-Model}s, and evaluation protocols. Together, they provide an essential resource for advancing audio-grounded role-playing research.
\end{abstract}

\section{Introduction}

\begin{figure}[h]
    \centering
    \includegraphics[width=1\linewidth]{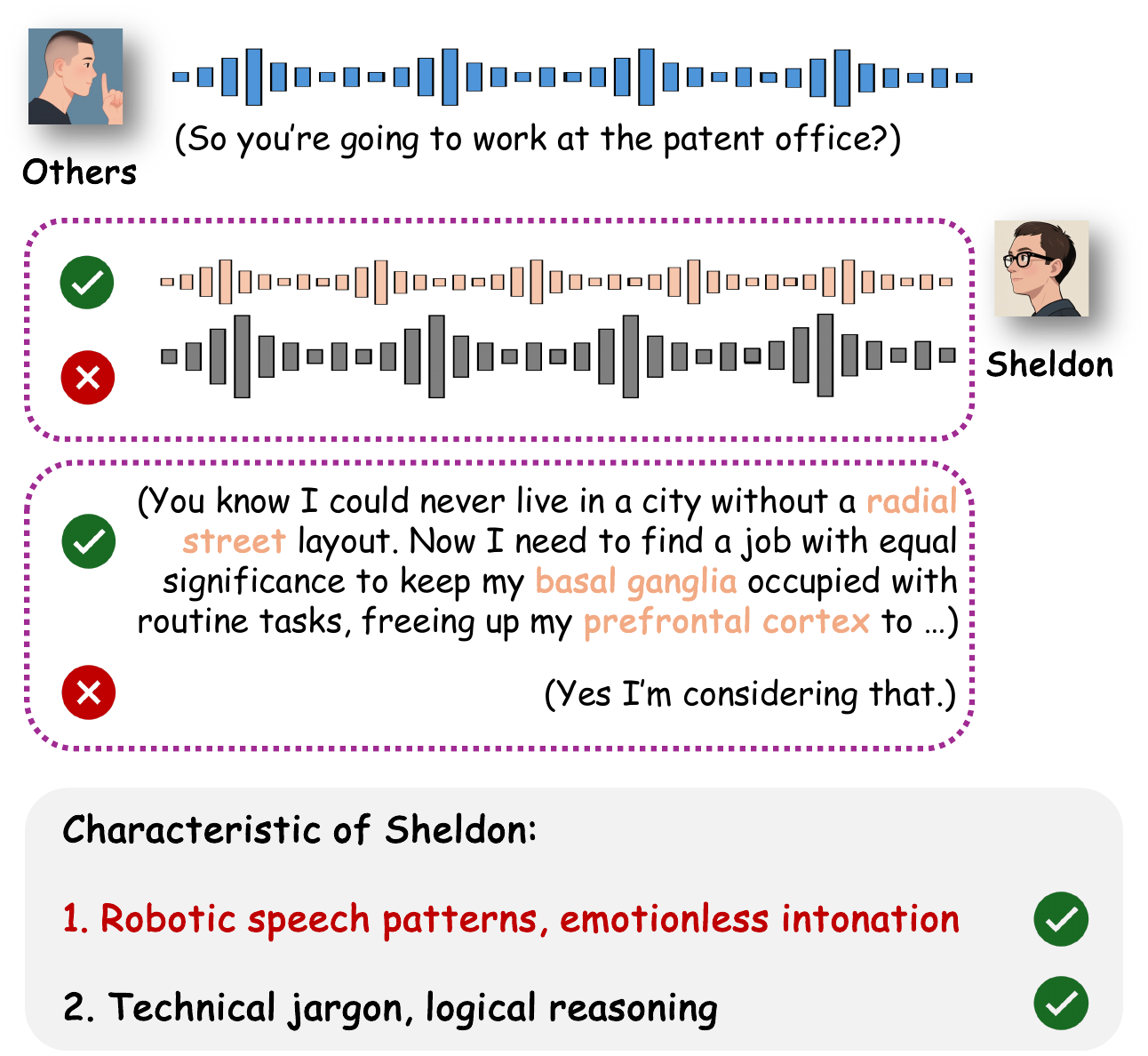}
    \caption{Audio Role-Playing case of Sheldon. The answer should satisfy not only the lexical similarity but also acoustic similarity.}
    \label{fig:sheldon}
\end{figure}
The evolution of role-playing capabilities in LLMs has revolutionized human-AI interaction paradigms. Contemporary systems demonstrate remarkable proficiency in textual persona simulation, serving as personalized assistants~\footnote{\href{https://chatgpt.com/}{https://chatgpt.com/}}, emotional companions\footnote{\href{https://replika.com/}{https://replika.com/}}, and social interaction proxies \cite{park2023generativeagentsinteractivesimulacra}. However, this progress remains fundamentally constrained by unimodal (text-only) interaction frameworks, which neglect the critical role of vocal characteristics in authentic character portrayal — a limitation that is particularly evident when simulating personas from audiovisual media, such as films and TV series.

Existing work on audio evaluation considers mainly content accuracy \cite{chen2024voicebenchbenchmarkingllmbasedvoice} \cite{hassid2024textuallypretrainedspeechlanguage} or only the conversational style \cite{liu2025vocalbenchbenchmarkingvocalconversational}. Vocal expression reflects an integral component of communication that varies considerably between individuals in different contexts \cite{cohen2015vocal}, and the acoustic properties of speech reflect key variables for understanding human behavior \cite{decety2006human}. Vocal expression is highly variable across individuals and contexts, and is influenced by several individual differences. Consider Dr. Sheldon Cooper from \textit{The Big Bang Theory} (Figure~\ref{fig:sheldon}): while text-based LLMs can mimic his technical jargon and logical reasoning, they fundamentally fail to capture his iconic robotic speech patterns — specifically, his faster-than-average speaking rate and emotionless intonation. This disconnect highlights a critical challenge in audio-grounded role-playing: authentic characterization requires not just \textit{what} the persona says, but precisely \textit{how} they say it — the synchronized alignment of semantic content and acoustic delivery that defines recognizable personalities.

To promote the solution of this challenge, we introduce \textbf{Audio Role-Playing (ARP)}, a novel task requiring dual-alignment of generated responses: (1) semantic consistency with character knowledge and speaking style, and (2) acoustic fidelity to vocal profiles. ARP task extends beyond speech synthesis — successful ARP systems must dynamically adapt both \textit{what} is said (content) and \textit{how} it's said (delivery) according to situational contexts and character traits.

Central to advancing ARP research is the creation of high-quality training data. Current attempts using general-purpose multi-modal models such as GPT-4o-Audio and MiniCPM-O-2.6, rely on prompt engineering for audio role-playing; however their zero-shot or one-shot performance remains suboptimal due to the absence of dedicated training corpora. Our solution, \textbf{AudioRole}, addresses the data scarcity through systematic curation of 13 TV series spanning 1K+ hours, which makes it support Audio Role-Playing for a wide range of characters.

The dataset's efficacy is validated through rigorous evaluation against state-of-the-art multi-modal models. When fine-tuned on AudioRole, evaluated by our dual-aspect evaluation framework, which assessed \textit{response quality} and \textit{role fidelity} for both acoustic and semantic. While keeping a good Acoustic Quality score, our ARP-Model achieves 0.31 Acoustic Personalization score, significantly outperforming the raw mode before fine-tuning, even outperform GPT-4o-Audio and MiniCPM-O-2.6. It also gets a Content Personalization score 0.36 higher than the raw model, which shows the ability of role-playing trained into the ARP-Model.

Our work has three key contributions: 
\begin{itemize}
    \item We formally define Audio Role-Playing as a dual-alignment task requiring synchronized generation of character-appropriate content (knowledge, speaking style) and acoustic properties (pitch, pacing, timbre).
    \item We construct AudioRole — the first large-scale dataset enabling systematic training of ARP systems, with 1M+ audio samples from more than 115 main characters, capturing nuanced vocal profiles across diverse TV characters.
    \item Experimental validation—corroborated by human perceptual studies—demonstrates that models trained on AudioRole achieve both higher Acoustic Personalization and higher Content Personalization score than the raw model before training. (Even higher than leading multi-modal models using zero-shot or one-shot prompting), proving the dataset's critical role in advancing audio-grounded role-playing.
\end{itemize}

To accelerate ARP research, we open-source a complete base comprising AudioRole, ARP-Eval, and 6 fine-tuned ARP-Models, which are potentially key resources for developing AI agents that truly embody digital personas.

\section{Related Work}
\paragraph{Role-Playing}
Recent advancements in role-playing with large language models (LLMs) have predominantly confined to textual modalities, overlooking the critical role of audio cues in enhancing role immersion and authenticity. Li et al. \cite{li2023chatharuhi} synthesized dialogues for 32 TV/animation characters using scripts and GPT-generated simulations, while Tu et al. \cite{tu2023characterchat} created 1,024 MBTI-based personas via ChatGPT-driven conversational agents. Chen et al. \cite{chen2024socialbenchsocialityevaluationroleplaying} proposed the first benchmark designed to systematically evaluate the sociality of role-playing conversational agents. The role-playing model's development emphasizes persona consistency through supervised fine-tuning (e.g., CharacterLLM \cite{shao2023character}) and in-context learning, with evaluation frameworks evolving from basic persona adherence to nuanced metrics\cite{tu2024charactereval}. Role-playing agents now emulate diverse personas—from fictional characters \cite{chen2023large} \cite{zhou2023characterglm} \cite{10.1145/3701551.3703583} \cite{10.1145/3589335.3651584} to user-specific clones \cite{li2021dialogue}  — to deliver emotional or sociological value \cite{gu2024agent}. Although \cite{zhan2025vstylebenchmarkvoicestyle} proposes a benchmark for voice style adaptation, it only treats role-playing as one of the four categories and directly scores it using an LLM, which omits the importance and complexity of the Audio Role-Playing task. \cite{jiang2025speechrolelargescaledatasetbenchmark} present a dataset of 98 roles and 112k conversations. In contrast to our work, their approach relies on GPT-4.1-2025-04-14 for dialogue generation and provides only one audio reference per character, which may compromise the authenticity and granularity needed for robust audio role-playing evaluation.

\paragraph{Voice Conversion}
Existing speech conversion research primarily focuses on acoustic transformation while preserving semantic content. Voice conversion (VC) techniques, such as AUTOVC \cite{qian2019autovc} and VAW-GAN \cite{hsu2017voice}, aim to modify speaker identity (e.g., accent, timbre) or emotional prosody through disentangled representations of content and style. Recent advancements like Text-guidedVC \cite{kuan2023towards} and HybridVC \cite{niu2024hybridvc} leverage text prompts or hybrid audio-text inputs to achieve flexible style transfer without parallel data. Voice cloning systems, including Tacotron-based models \cite{zhao2020research} and NAUTILUS \cite{luong2020nautilus}, synthesize speech in target voices by replicating vocal traits while retaining input text content. These works uniformly prioritize content preservation, modifying acoustic attributes without altering semantic meaning.

In contrast, our work introduces audio-grounded role-playing, which diverges fundamentally by jointly transforming both acoustic features and semantic content to align with specific character personas. While traditional VC ensures content consistency (e.g., "Hello" remains "Hello" across styles), our framework enables role-specific responses (e.g., a medieval knight might reply, "Hark! Who goes there?" instead of "Hello") and even refusal to engage (e.g., a stoic guard ignoring casual queries). This dual focus on acoustic-semantic role alignment bridges the gap between speech conversion and LLM-based role-playing, where character authenticity requires synchronized adaptation of acoustic style and semantic content—a paradigm unexplored in prior speech conversion research.

\section{Audio Role-Playing Task}
We formally define the Audio Role-Playing task as follows: Given a target character $C$, operationally defined as a set of reference audio samples exhibiting the character's unique attributes, and an input audio $X_a$, synthesize an output audio response $X_b$ that simultaneously satisfies: which satisfies both: (1) The semantic content of $X_b$ demonstrates contextual appropriateness while manifesting $C$'s distinctive. (2) The vocal output $X_b$ faithfully preserves $C$'s acoustic identity.
\subsection{AudioRole Construction}
\begin{figure*}
    \centering
    \includegraphics[width=1\linewidth]{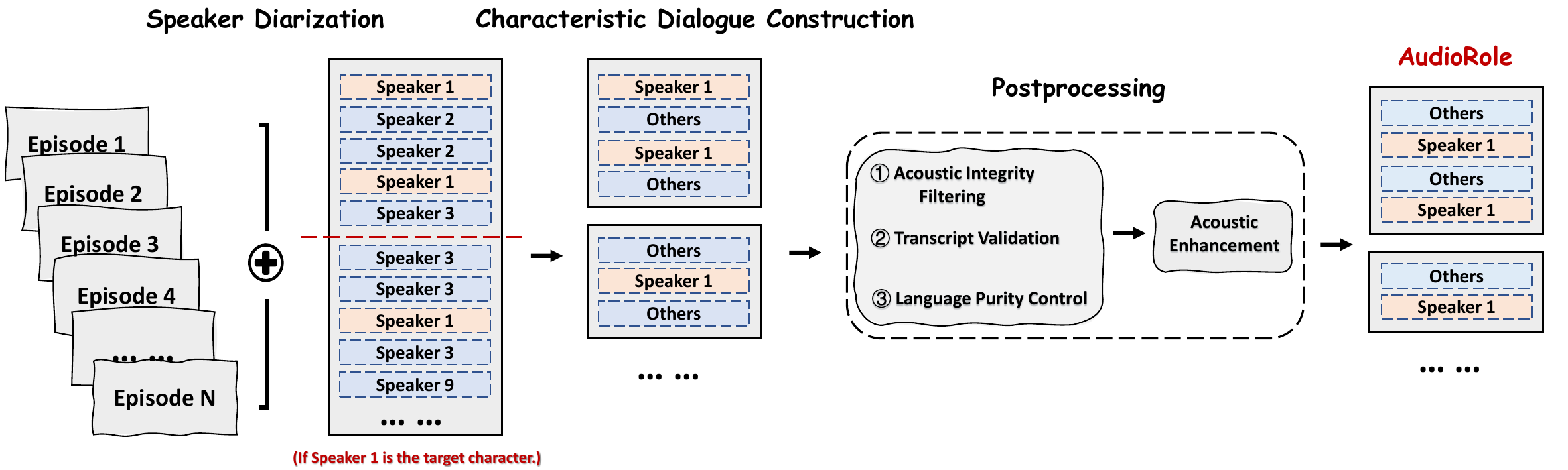}
    \caption{The pipeline of AudioRole Construction.}
    \label{fig:pipeline}
\end{figure*}

We aim to ensure vocal consistency for each specific character; therefore, we focus on long-running TV series where main characters maintain stable vocal characteristics across seasons. We avoid series with potential actor changes, age-related voice variations, or insufficient dialogue. Consequently, we have selected TV series featured in a recent dataset called ``Bazinga'' \cite{lerner-etal-2022-bazinga}\footnote{ We used \href{https://huggingface.co/datasets/bazinga/bazinga}{``Bazinga!''} exclusively for the purposes of this research. If you wish to use our dataset, please ensure to follow the copyright guidelines associated with ``Bazinga!''.} which contains 16 TV series in the language of English. Our dataset construction pipeline consists of three core phases: speaker diarization, context-aware dialogue extraction, and postprocessing. These phases are illustrated in Figure \ref{fig:pipeline}.
\subsubsection{Speaker Diarization}
For each TV series, we aggregate all episode audios into continuous streams using FFmpeg's\footnote{\href{https://github.com/FFmpeg/FFmpeg}{https://github.com/FFmpeg/FFmpeg}} lossless waveform concatenation processing. This preserves original 16kHz sampling rates while normalizing audio codecs into uncompressed WAV format. 

Pyannote 3.1\footnote{\href{https://huggingface.co/pyannote/speaker-diarization-3.1}{https://huggingface.co/pyannote/speaker-diarization-3.1}} is employed for diarization to extract speaker timestamps, which generate RTTM (Rich Transcription Time Marked) files containing precise speaker segments with temporal resolution better than 100 milliseconds.

The subsequent audio segmentation phase extracts speaker-specific clips using PyTorch's tensor operations. By calculating exact frame positions from RTTM timestamps and sample rates, we precisely slice the concatenated audio stream while maintaining original quality. These segments are then merged per speaker identity through PyTorch's tensor concatenation.
The output contains the complete dialogue collection of each character that is used to choose the target character.

\subsubsection{Characteristic Dialogue Construction}
In our dataset, we only apply the following steps for six chosen TV series, which we called AudioRole-Demo, and if you want to make your own character dataset, you can just follow our procedure. We simply take the one main character in each TV series to be the target character, but we also support building any character that is in these TV series. Building upon the merged audio, we sort the audio files and choose the largest one to be the audio of the main character. To ensure validity, we manually verify the candidates per TV series, correcting incorrect-character or non-character audio segments.

Dialogue scene extraction then partitions the continuous audio stream into conversational units using temporal dynamics. We define a dialogue scene as a sequence where: 1) multiple speakers participate, 2) pauses between utterances never exceed 3s\footnote{We tested segmentation thresholds of 2s, 3s, and 4s on a control set, the 3s threshold provided the optimal balance for preserving complete dialogue turns while accurately segmenting distinct interactions.}, and 3) contains at least one utterance from the target character.

Each validated scene undergoes role-centric restructuring to form stimulus-response pairs. Consecutive utterances from non-target speakers merge into unified "Speaker A" and consecutive utterances from the target speaker merge into unified "Speaker B", while preserving the temporal order of all characters' responses. 

\subsubsection{Postprocessing}
To establish a high-quality multimodal corpus, we implement a rigorous postprocessing pipeline ensuring synchronization between audio waveforms and textual transcripts. The transcription generation employs Moore Threads' MooER model\footnote{\href{https://github.com/MooreThreads/MooER}{https://github.com/MooreThreads/MooER}}, a LLM-based speech recognition and translation model.

Four quality control measures are systematically applied in order to purify by removing fragments of questionable quality:
\paragraph{Acoustic Integrity Filtering}Discard audio segments shorter than 1s and truncate those exceeding 30s to focus on meaningful utterances, removing samples with abrupt cuts or incomplete phrases.
\paragraph{Transcript Validation}For there may be transcription questions, always generating very long or very short texts, we eliminate transcripts with fewer than 2 characters or exceeding 512 characters\footnote{Our manual inspection revealed that over 90\% of ASR failures manifested as either hallucinated infinite repetitions or cross-lingual confusion. Setting the length limit and the following language purity threshold proved to be an effective method.}. 
\paragraph{Language Purity Control}For the transcription generation, sometimes transcribe English audio to Chinese audio or Chinese audio to English audio. Take English audio as an example, we implement cross-lingual filtering by discarding segments where the non-target language character ratio exceeds 0.1."

\paragraph{Acoustic Enhancement}We use Deep Filter Net\footnote{\href{https://github.com/Rikorose/DeepFilterNet}{https://github.com/Rikorose/DeepFilterNet}}, a low-complexity speech enhancement framework designed for full-band audio based on deep filtering technology to suppress audio’s noise.

Finally, we got the AudioRole dataset and AudioRole-Demo dataset, which we will all release.
\subsection{Dataset Statistics}
Our dataset contains 515.66h of audio generated from 13 TV series containing 1M+ character-grounded dialogues. As shown in Table \ref{tab:AudioRole}, TV series like \textit{The Walking Dead} have more than 75\% loss when extracting speaking audios from the original audio, showing that the zombie TV series uses a lot of ambient sound, silence, and visual storytelling instead of dense dialogue. In contrast, sitcoms or pseudo-documentary style TV series like \textit{Friends}, \textit{The Big Bang Theory}, and \textit{The Office} rely on dense dialogues to create laughs and advance the plot, and maintain the comedy rhythm and audience entertainment experience through rapid line interactions in fixed scenes, so the environmental sound loss is less than 40\%. 

All the 515.66h audio contains more than 115 main characters' multi-turn dialogue, which makes it a rich dataset with the potential to build a role-playing dataset.

\begin{table}
  \caption{Statistic for AudioRole. Speech time is the time with the character speaking, which is generated from the Audio time. All time units are hours.}
  \label{tab:AudioRole}
  \setlength{\tabcolsep}{1.8pt}
  \small
  \begin{tabular}{lcccc}
    \toprule
    TV series&Episode&Speech time&Audio time\\
    \midrule
      24& 192 &56.25& 134.41 \\
      Battlestar Galactica& 71 & 19.64 & 52.28\\
      Breaking Bad& 61 & 17.16 & 46.49\\
      Buffy the Vampire Slayer& 143 & 43.42& 101.32\\
      ER& 330 & 135.41 & 235.49\\
      Friends& 233 & 51.32 & 84.93 \\
      Game of Thrones& 60 & 18.45 & 53.17\\
      Homeland& 70 & 19.87 & 57.83\\
      Lost& 104 & 22.72 & 74.61 \\
      Six Feet Under& 63 & 25.14 & 56.72\\
      The Big Bang Theory& 207 & 44.17 & 68.69\\
      The Office& 188 & 44.65 & 71.76\\
      The Walking Dead& 99 & 17.46 & 72.17\\
  \bottomrule
\end{tabular}
\end{table}

The end-to-end training is computationally intensive, which constrained us from training models for every character in this initial AudioRole. So following the pipeline introduced above, we built the role-playing dataset for six main characters from six different TV series, which we named AudioRole-Demo, containing six main characters, who are Jack Bauer from \textit{24}, Laura Roslin from \textit{Battlestar Galactica}, Walter White from \textit{Breaking Bad}, Buffy from \textit{Buffy The Vampire Slayer}, Tyrion Lannister from \textit{Game Of Thrones}, and Sheldon from \textit{The Big Bang Theory}. 

The statistic is shown in Table \ref{tab:AudioRole-Demo}. We split all the multi-turn dialogues in the training set into single-turn dialogues and directly appended them to the original multi-turn dialogues. Our AudioRole-Demo covers the speaking times range from 1.50h to 12.5h and characters from a hard-core agent who speaks briefly, directly and commandingly to a verbose, jargon-filled, socially awkward, outspoken geek, which covers a wide range of speaking time and a wide range of characters, making it fully demonstrating our datasets' high quality through the experiment below.

\begin{table}
  \caption{Statistic for AudioRole-Demo. ``Scenes'' are before the postprocessing step, and all time units are hours.}
  \label{tab:AudioRole-Demo}
  \setlength{\tabcolsep}{3.7pt}
  \small
  \begin{tabular}{lcccc}
    \toprule
    Characters&Times&Scenes&Turns-train&Turns-test\\
    \midrule
      Jack Bauer& 5.30 & 2624 & 2075 & 90\\
      Laura Roslin& 1.50 & 1117 & 600 & 42\\
      Walter White& 2.76 & 994 & 682 & 30\\
      Buffy& 7.70 & 1703 & 3589 & 190\\
      Tyrion Lannister& 1.34 & 1300 & 339 & 22\\
      Sheldon& 12.50 & 723 & 4209 & 250\\
  \bottomrule
\end{tabular}
\end{table}

To verify the representativeness of AudioRole-Demo, we conducted a comparative analysis on 100 randomly sampled clips from the full corpus versus the demo subset. The results demonstrate high consistency across key metrics, with the full dataset and demo subset achieving comparable Average SNR (19.37 dB vs. 18.24 dB), Average Noise Duration (0.69s vs. 0.73s), and Speaker Overlap Rate (8\% vs. 11\%). These aligned figures confirm that the evaluated subset accurately reflects the broader corpus quality, validating the effective generalization of our pipeline.

\subsection{ARP-Model}
To demonstrate the high quality of our dataset, we trained the ARP-Model using the AudioRole-Demo. We use the GLM-4-Voice \cite{zeng2024glm} as the base model, which is an end-to-end voice model that can directly understand and generate Chinese and English speech. It has three models: (1) GLM-4-Voice-Tokenizer takes audio input and converts raw audio input into discrete tokens. (2) GLM-4-Voice-9B takes the token inputs, then thinks and responds also in discrete tokens. (3) GLM-4-Voice-Decoder converts discrete speech tokens into continuous speech outputs.

To make the model have the ability to role-play the character in both semantic content and acoustic identity, we trained the GLM-4-Voice-9B model and GLM-4-Voice-Decoder separately.

To train the GLM-4-Voice-9B model, we formatted the training data into 13 text tokens and then 26 audio tokens for each character in our AudioRole-Demo dataset, and then followed the common training procedure of Colossal AI platform\footnote{\href{https://github.com/hpcaitech/ColossalAI}{https://github.com/hpcaitech/ColossalAI}} with 20 training epochs. Considering the GLM-4-Voice-Decoder, we first extract all the audios of the target characters to form six high-quality unsupervised speech data from a single speaker, then trained the flow matching model of GLM-4-Voice-Decoder from scratch about 15 epochs.

We do the training on 8$\times$40GB GPUs about 30 hours in total. Finally, we directly follow the original structure of GLM-4-Voice, and put the untrained tokenizer and the two trained models together in order for each character to form our 6 ARP-Models.

\subsection{ARP-Eval}
The dual-alignment paradigm of Audio Role-Playing necessitates a comprehensive evaluation framework that simultaneously addresses both acoustic and semantic dimensions of character embodiment. We propose ARP-Eval as a unified assessment framework that rigorously quantifies 4 critical aspects of performance. This framework emerges from the fundamental observation that authentic character portrayal requires not only high-quality speech synthesis but also consistent preservation of character-specific attributes across multiple modalities.

\paragraph{Acoustic Quality (AQ)} AQ establishes the basic requirement for perceptual acceptability. We employ Audiobox's pre-trained aesthetic scoring model \cite{tjandra2025metaaudioboxaestheticsunified}\footnote{\href{https://github.com/facebookresearch/audiobox-aesthetics}{https://github.com/facebookresearch/audiobox-aesthetics}} to compute production quality scores in a range of 0 to 10, quantifying technical attributes including signal-to-noise ratio, harmonic-to-noise ratio, and spectral flatness. This ensures $X_b$ meets broadcast-standard technical criteria before character-specific evaluation.

\paragraph{Content Quality (CQ)} CQ requires output audio $X_b$ to demonstrate both contextual appropriateness and domain accuracy while maintaining the target character $C$'s persona. For instance, when input audio $X_a$ queries ``Explain quantum entanglement,'' $X_b$ from $C$-Sheldon should respond: \textit{``While quantum entanglement appears spooky, it's simply correlated quantum states persisting after particle separation''} -- employing scientific lexicon without deviating into unrelated topics. Our pipeline using whisper-turbo\footnote{\href{https://github.com/openai/whisper}{https://github.com/openai/whisper}} transcribes $X_b$ to text $T_b$, then computes semantic alignment using the GPT-4o model, and the score is in a range of 0 to 2. The prompt used is shown in Table \ref{tab:prompt_cq}.

\paragraph{Acoustic Personalization (AP)} AP quantifies voice characteristic preservation through PyAnnotate's speaker embedding model\footnote{\href{https://huggingface.co/pyannote/wespeaker-voxceleb-resnet34-LM}{https://huggingface.co/pyannote/wespeaker-voxceleb-resnet34-LM}}. Given reference audio samples $X_c$ of $C$ and synthesized $X_b$, we extract their embeddings $e_c$ and $e_b$ and then calculate AP using cosine similarity.  
 The score ranges from -1 to 1, where higher values indicate greater similarity to the target character's acoustic identity.

\paragraph{Content Personalization (CP)} CP evaluates stylistic consistency using GPT-4o-audio multi-modal reasoning ability. The model takes in $X_b$ and a reference audio sample $X_c$, and a prompt to analyze whether they show the same character's style in a range of 0 to 2. The prompt used is shown in Table \ref{tab:prompt_cp}.

By combining objective signal measurements with learned character representations, ARP-Eval provides comprehensive insights into system performance, while this multifaceted approach effectively captures the complex interplay between speech quality, contextual intelligence, and character consistency, which defines successful audio role-playing implementations.

\section{Experiment}

\subsection{Baselines}
To establish a comprehensive benchmark for the Audio Role-Playing (ARP) task, we select the following state-of-the-art models as baselines:
\begin{itemize}
\item \textbf{GPT-4o Audio\footnote{\href{https://platform.openai.com/docs/models/gpt-4o-audio-preview}{https://platform.openai.com/docs/models/gpt-4o-audio-preview}}:} As one of the most powerful proprietary multimodal models, GPT-4o Audio sets a strong baseline for general audio understanding and generation, and for our evaluations, we utilize its default ``alloy'' voice profile. This baseline represents the zero-shot capability of a top-tier commercial system for our task.

\item \textbf{MiniCPM-o 2.6\footnote{\href{https://openbmb.notion.site/MiniCPM-o-2-6-GPT-4o-188ede1b7a558084b3aedd669cb80730}{https://openbmb.notion.site/MiniCPM-o-2-6}}:} This open-source model is a strong contender that explicitly supports voice style adaptation and one-shot role-playing, making it a highly relevant and strong baseline for assessing few-shot character adaptation performance without specialized training on our dataset.

\item \textbf{GLM-4-Voice:} We use the original, pretrained GLM-4-Voice model as our base model. It allows us to isolate and quantify the performance improvement gained specifically from fine-tuning on our proposed AudioRole dataset, separate from any inherent capabilities of the underlying architecture.

\end{itemize}

For our fine-tuned ARP-Model, the model requires only the user query, as the persona is intrinsic to the model weights. For baselines, we utilized a direct prompt as shown in Table \ref{tab:prompt_baseline} to test their inherent one-shot role-playing capabilities, which provides a fair baseline comparison without heavy prompt engineering


\subsection{Experiment Analysis}

We conduct a comprehensive evaluation of the ARP-Model against baseline models on the AudioRole-demo dataset. As shown in Table \ref{tab:experiment}, our experiment reveals three key findings, including both quantitative metrics and qualitative observations:
\begin{table*}
  \caption{Main experimental results. For each metric (higher is better) in ``Avg.'', the highest score is highlighted in \underline{\textbf{bold and underline}}, while the second best is marked with \underline{\textit{italics and underline}}. All values retain two decimal places.}
  \label{tab:experiment}
  \setlength{\tabcolsep}{4.15pt}
  \small
  \begin{tabular}{lcccccccccccccccc}
    \toprule
    Characters&\multicolumn{4}{c}{GPT-4o Audio}&\multicolumn{4}{|c|}{MiniCPM-o 2.6}&\multicolumn{4}{|c|}{GLM-4-Voice}&\multicolumn{4}{|c}{ARP-Model}\\
     &AQ&CQ&AP&CP&AQ&CQ&AP&CP&AQ&CQ&AP&CP&AQ&CQ&AP&CP\\
     \midrule
     Jack Bauer & 7.80&1.90 &0.02 &1.30  &6.50 & 0.66& 0.13&0.30 &7.50 &0.37 &0.05 & 0.24& 6.00& 0.22&0.23 & 0.49\\
     Laura Roslin  & 7.70&1.80&-0.02&0.80&7.10&0.71&0.24&0.41&7.50&0.50& 0.03&0.24&6.20 &0.33 &0.33 & 0.41\\
     Walter White &7.50 &1.60 & 0.08& 1.30& 7.10&0.70 &0.22& 0.38& 7.60&0.37 & 0.07&0.43&6.40 & 0.50&0.26 &0.17 \\
     Buffy & 7.70& 1.80& 0.01&1.10 & 6.80&0.69 &0.09 & 0.34& 7.60& 0.52&0.10 &0.25 &6.70 &0.43 &0.39 &0.36 \\
     Tyrion Lannister &7.60 &1.90 &0.00 & 0.95 &6.90 &0.59 & 0.14&0.14 & 7.60&0.36 &0.04 & 0.10&  6.8& 0.55&0.25 &0.26 \\
     Sheldon &7.70 &1.30&0.29 &1.00 & 6.80& 1.10& 0.24& 0.54&7.60 &0.43 & 0.01& 0.32&6.70 & 0.28&0.42 &0.44 \\
     \midrule
     Avg. &\underline{\textbf{7.70}} &\underline{\textbf{1.70}} &0.06 &\underline{\textbf{1.10}} &6.90 & \underline{\textit{0.74}}&\underline{\textit{0.17}} &0.35 &\underline{\textit{7.60}} & 0.43& 0.05& 0.26& 6.50&0.39 &\underline{\textbf{0.31}}&\underline{\textit{0.36}} \\
  \bottomrule
\end{tabular}
\end{table*}

\paragraph{Acoustic Content Tradeoff} The experimental results show that although the audio quality is reduced to a certain extent (ARP-Model AQ=6.5 vs GLM-4-Voice=7.6), our model showcases persona adaptation without catastrophic forgetting of fundamental speech capabilities. This controlled degradation primarily stems from two technical factors: (1) the inherent conflict between character-specific vocal patterns (e.g., Sheldon's accelerated speech rate) and the base model's default prosodic templates during fine-tuning; (2) preservation of environmental noise from TV recordings (outputs from our AudioRole-demo dataset, which achieve AQ=6.3 under the same evaluation)  prevents smoothing to synthetic speech patterns. Notably, GPT-4o Audio's anomalously high CQ scores may reveal evaluation bias - when using GPT-4 as both generator and evaluator, the self-consistency preference inflates scores.

\paragraph{Acoustic Personalization Superiority} The fidelity of the role-playing experience is most sensitive to Acoustic Personalization. In contrast to the more tolerable variations in AQ, CQ, and CP, differences in AP have a pronounced effect on the perceived authenticity of a character. ARP-Model demonstrates 5× higher AP scores than GPT-4o Audio and 80\% improvement over MiniCPM-O-2.6, validating the necessity of dedicated role-specific training. The remaining gap from score 1.0 primarily occurs in three scenarios: 1) The outputs from AudioRole-demo (the reference answer) has a different answer strategy from the ARP-Model, resulting in differences in speaking speed, emotional fluctuations, etc., 2) The ARP-Model's answer still has some noise or somewhere badly learned after fine-tuning, and 3) The context of the non-linear nature of high-dimensional cosine similarity.

\paragraph{Content Personalization Dynamics} Despite lower absolute CQ scores, ARP-Model's CP outperforms GLM-4-Voice by 38\% and even a little higher than MiniCPM-O-2.6's one-shot performance. This indicates successful injection of character-typical speech patterns. When meeting the same input, the ARP-Model can generate more personalized output compared to the raw GLM-4-Voice model. However, the CP gap between ARP-Model and the reference answer exposes limitations in contextual adaptation - the model occasionally generates character-consistent but situationally inappropriate responses (e.g., using humorous in solemn contexts).
\begin{figure}[ht]
    \centering
    \includegraphics[width=1\linewidth]{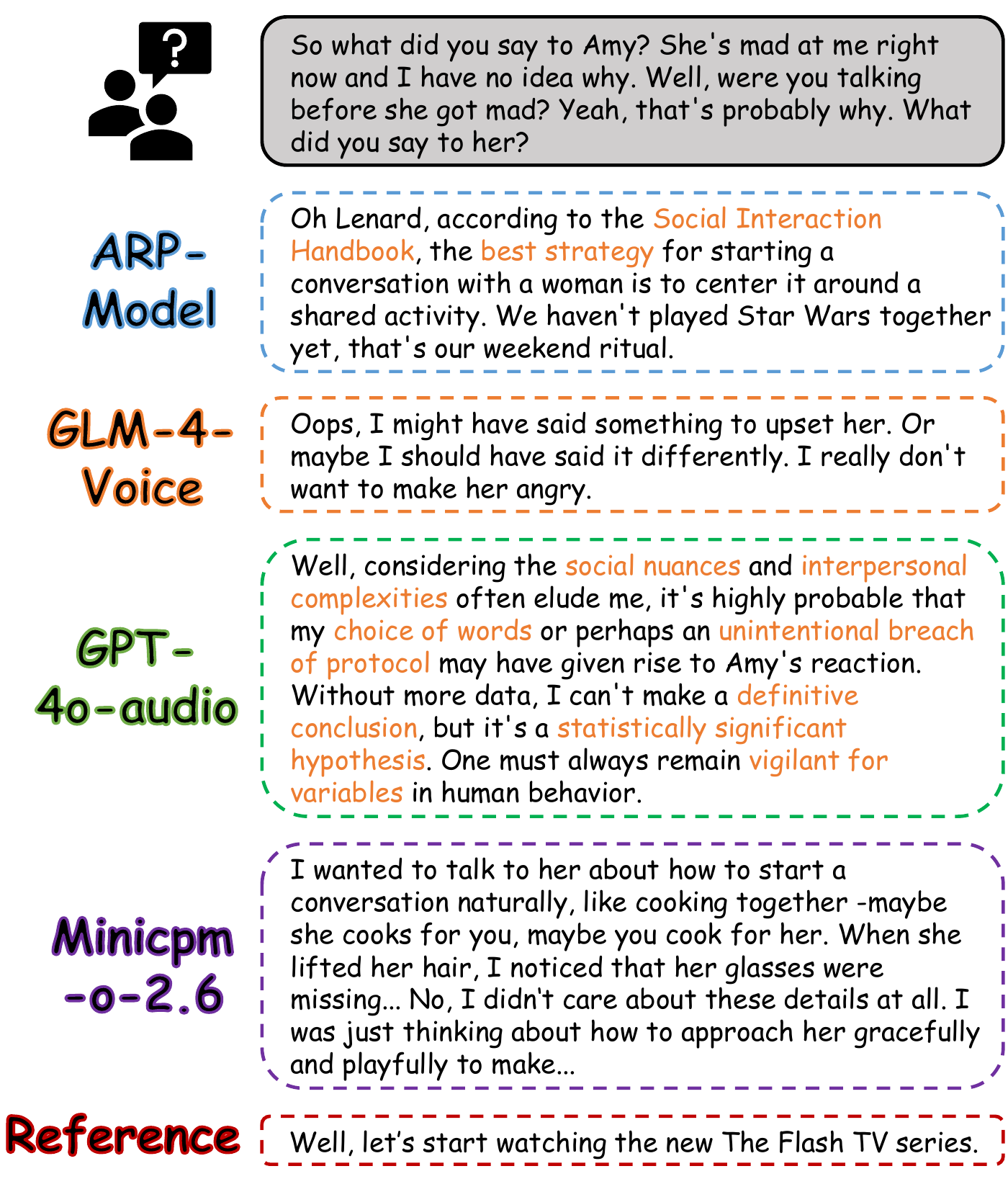}
    \caption{One typical case of Sheldon, and the words in orange show the scientific terms used by the models to imitate Sheldon. ``Reference'' means the output from the AudioRole-Demo dataset}.
    \label{fig:case}
\end{figure}

Figure \ref{fig:case} shows a representative example of how each model responds to a query directed at Sheldon. The ARP-Model’s answer clearly reflects his persona: it uses precise scientific terms, adopts a condescending tone, and notably avoids any apology or show of emotion. When it encounters discomfort, the ARP-Model even changes the subject abruptly (e.g., redirecting the conversation to a physics concept) – a classic Sheldon move. The reference answer from the TV show behaves very similarly, confirming that the ARP-Model has captured many of these quirks. In contrast, the raw GLM-4-Voice response, while factually reasonable, misses the mark on character. It might apologize or express empathy, which Sheldon would never do. The GPT-4o Audio output is fluent and technically detailed, but it actually overuses jargon; it sounds as if the model is deliberately forcing scientific buzzwords, making the response feel contrived. Essentially, GPT-4o’s answer is too polished and verbose – akin to an actor trying too hard. Finally, the MiniCPM-2.6 answer is longer and more emotional, injecting sympathy or personalization, which again does not fit Sheldon’s unemotional style. 

This qualitative case underscores our quantitative findings. The ARP-Model best approximates the true character by naturally blending persona and content: its answer may have slightly lower audio clarity, but it ``feels'' like Sheldon. The other models either sound too polite, too artificially scholarly, or too sentimental. In sum, the targeted fine-tuning of the ARP-Model leads to more authentic role-playing speech than these generic approaches.

\subsection{Human Perceptual Evaluation}
While automated metrics provide a scalable assessment, human perception remains the gold standard for evaluating role-playing authenticity. To validate our automatic metrics (AP and CP), we conducted a blind human evaluation on 50 randomly selected test samples. Two annotators ranked the responses of four models (GPT-4o, MiniCPM-o-2.6, GLM-4-Voice, and our ARP-Model) from 1 (worst) to 4 (best) based on Acoustic Personalization and Content Personalization as shown in Table \ref{tab:Human-Eval}.

The results strongly validate our automatic metrics, and both annotators said ARP-Model's voice is significantly better than others, which is almost the same as the character. The high inter-annotator agreement (>76\%) further confirms the reliability of these findings.
\begin{table}
  \caption{Human perceptual evaluation results. The two Rank scores are the ``Average scores'' and ``Agree.'' represents ``Agreement''.}
  \label{tab:Human-Eval}
  \setlength{\tabcolsep}{2.2pt}
  \small
  \begin{tabular}{lcccc}
    \toprule
    Model&AP Rank&CP Rank&AP Agree.&CP Agree.\\
    \midrule
      ARP-Model& 4.00 & 2.75 & 100\% & 90\%\\
      GPT-4o-Audio& 1.61 & 3.58 & 84\% & 76\%\\
      MiniCPM-o-2.6& 3.00 & 2.56 & 100\% & 92\%\\
      GLM-4-Voice& 1.39 & 1.11 & 84\% & 88\%\\
  \bottomrule
\end{tabular}
\end{table}

\section{Conclusion}
We present AudioRole, a novel framework for audio-grounded character role-playing that bridges text-based persona simulation and vocal identity preservation. We release a large-scale dataset AudioRole of 515+ hours from 13 TV series with 1M+ character-aligned audio-text pairs and a human-checked AudioRole-Demo, which contains 6 main characters' multi-turn dialogue data over 31h. By both automated metrics from ARP-Eval and human listeners, the experimental results demonstrate that our ARP-Model successfully combines character-specific speech patterns with appropriate semantic content. 

This work provides essential resources for developing authentic multi-modal AI personas to promote the development in this area, and you can easily create your own favorite character's dataset just following our pipeline.

\section*{Ethical considerations}
AudioRole is derived from a publicly available dataset consisting of audio from popular TV series. We believe this research uses publicly accessible, fictional content for non-commercial purposes presents no major ethical concerns.

\section*{Limitations}
Despite implementing advanced diarization techniques and noise-suppression methodologies to ensure dataset integrity, speaker attribution errors continue to occur. Notably, in scenes with conversational overlap, non-stationary environmental noise can still compromise precise voiceprint extraction in certain acoustic conditions.

Current assessment protocols, while comprehensive, exhibit limitations in quantifying nuanced temporal speech dynamics and fully capturing cross-modal alignment, particularly the synergistic relationship between vocal delivery and semantic content during character portrayal. 

Our current experimental setup focuses on single-turn evaluation. While multi-turn consistency is vital for long-term interaction, we prioritized isolating the acoustic role-playing capability—a foundational step before addressing complex contextual memory. However, it also notes that the AudioRole dataset itself preserves full dialogue history and speaker turns, making it natively ready for future research into multi-turn consistency and persona stability.

These identified constraints underscore critical pathways for future refinement in developing robust multi-modal role-playing systems. Addressing these limitations—through expanded data diversity, enhanced audio processing pipelines, refined evaluation metrics, and multi-turn testing — will be pivotal for advancing the field.


\bibliography{custom}

@misc{park2023generativeagentsinteractivesimulacra,
      title={Generative Agents: Interactive Simulacra of Human Behavior}, 
      author={Joon Sung Park and Joseph C. O'Brien and Carrie J. Cai and Meredith Ringel Morris and Percy Liang and Michael S. Bernstein},
      year={2023},
      eprint={2304.03442},
      archivePrefix={arXiv},
      primaryClass={cs.HC},
      url={https://arxiv.org/abs/2304.03442}, 
}

@article{cohen2015vocal,
  title={Vocal acoustic analysis as a biometric indicator of information processing: Implications for neurological and psychiatric disorders},
  author={Cohen, Alex S and Dinzeo, Thomas J and Donovan, Neila J and Brown, Caitlin E and Morrison, Sean C},
  journal={Psychiatry Research},
  volume={226},
  number={1},
  pages={235--241},
  year={2015},
  publisher={Elsevier}
}

@article{decety2006human,
  title={Human empathy through the lens of social neuroscience},
  author={Decety, Jean and Lamm, Claus},
  journal={The scientific World journal},
  volume={6},
  number={1},
  pages={1146--1163},
  year={2006},
  publisher={Wiley Online Library}
}

@article{li2023chatharuhi,
  title={Chatharuhi: Reviving anime character in reality via large language model},
  author={Li, Cheng and Leng, Ziang and Yan, Chenxi and Shen, Junyi and Wang, Hao and Mi, Weishi and Fei, Yaying and Feng, Xiaoyang and Yan, Song and Wang, HaoSheng and others},
  journal={arXiv preprint arXiv:2308.09597},
  year={2023}
}

@article{tu2023characterchat,
  title={Characterchat: Learning towards conversational ai with personalized social support},
  author={Tu, Quan and Chen, Chuanqi and Li, Jinpeng and Li, Yanran and Shang, Shuo and Zhao, Dongyan and Wang, Ran and Yan, Rui},
  journal={arXiv preprint arXiv:2308.10278},
  year={2023}
}

@article{shao2023character,
  title={Character-llm: A trainable agent for role-playing},
  author={Shao, Yunfan and Li, Linyang and Dai, Junqi and Qiu, Xipeng},
  journal={arXiv preprint arXiv:2310.10158},
  year={2023}
}

@article{tu2024charactereval,
  title={Charactereval: A chinese benchmark for role-playing conversational agent evaluation},
  author={Tu, Quan and Fan, Shilong and Tian, Zihang and Yan, Rui},
  journal={arXiv preprint arXiv:2401.01275},
  year={2024}
}

@inproceedings{chen2023large,
  title={Large language models meet harry potter: A dataset for aligning dialogue agents with characters},
  author={Chen, Nuo and Wang, Yan and Jiang, Haiyun and Cai, Deng and Li, Yuhan and Chen, Ziyang and Wang, Longyue and Li, Jia},
  booktitle={Findings of the Association for Computational Linguistics: EMNLP 2023},
  pages={8506--8520},
  year={2023}
}

@article{zhou2023characterglm,
  title={Characterglm: Customizing chinese conversational ai characters with large language models},
  author={Zhou, Jinfeng and Chen, Zhuang and Wan, Dazhen and Wen, Bosi and Song, Yi and Yu, Jifan and Huang, Yongkang and Peng, Libiao and Yang, Jiaming and Xiao, Xiyao and others},
  journal={arXiv preprint arXiv:2311.16832},
  year={2023}
}

@article{li2021dialogue,
  title={Dialogue history matters! personalized response selection in multi-turn retrieval-based chatbots},
  author={Li, Juntao and Liu, Chang and Tao, Chongyang and Chan, Zhangming and Zhao, Dongyan and Zhang, Min and Yan, Rui},
  journal={ACM Transactions on Information Systems (TOIS)},
  volume={39},
  number={4},
  pages={1--25},
  year={2021},
  publisher={ACM New York, NY}
}

@article{gu2024agent,
  title={Agent group chat: An interactive group chat simulacra for better eliciting collective emergent behavior},
  author={Gu, Zhouhong and Zhu, Xiaoxuan and Guo, Haoran and Zhang, Lin and Cai, Yin and Shen, Hao and Chen, Jiangjie and Ye, Zheyu and Dai, Yifei and Gao, Yan and others},
  journal={arXiv e-prints},
  pages={arXiv--2403},
  year={2024}
}

@inproceedings{qian2019autovc,
  title={Autovc: Zero-shot voice style transfer with only autoencoder loss},
  author={Qian, Kaizhi and Zhang, Yang and Chang, Shiyu and Yang, Xuesong and Hasegawa-Johnson, Mark},
  booktitle={International Conference on Machine Learning},
  pages={5210--5219},
  year={2019},
  organization={PMLR}
}

@article{hsu2017voice,
  title={Voice conversion from unaligned corpora using variational autoencoding wasserstein generative adversarial networks},
  author={Hsu, Chin-Cheng and Hwang, Hsin-Te and Wu, Yi-Chiao and Tsao, Yu and Wang, Hsin-Min},
  journal={arXiv preprint arXiv:1704.00849},
  year={2017}
}

@article{zeng2024glm,
  title={Glm-4-voice: Towards intelligent and human-like end-to-end spoken chatbot},
  author={Zeng, Aohan and Du, Zhengxiao and Liu, Mingdao and Wang, Kedong and Jiang, Shengmin and Zhao, Lei and Dong, Yuxiao and Tang, Jie},
  journal={arXiv preprint arXiv:2412.02612},
  year={2024}
}

@inproceedings{kuan2023towards,
  title={Towards general-purpose text-instruction-guided voice conversion},
  author={Kuan, Chun-Yi and Li, Chen-An and Hsu, Tsu-Yuan and Lin, Tse-Yang and Chung, Ho-Lam and Chang, Kai-Wei and Chang, Shuo-Yiin and Lee, Hung-yi},
  booktitle={2023 IEEE Automatic Speech Recognition and Understanding Workshop (ASRU)},
  pages={1--8},
  year={2023},
  organization={IEEE}
}

@article{niu2024hybridvc,
  title={Hybridvc: Efficient voice style conversion with text and audio prompts},
  author={Niu, Xinlei and Zhang, Jing and Martin, Charles Patrick},
  journal={arXiv preprint arXiv:2404.15637},
  year={2024}
}

@inproceedings{zhao2020research,
  title={Research on voice cloning with a few samples},
  author={Zhao, Li and Chen, Feifan},
  booktitle={2020 International Conference on Computer Network, Electronic and Automation (ICCNEA)},
  pages={323--328},
  year={2020},
  organization={IEEE}
}

@article{luong2020nautilus,
  title={Nautilus: a versatile voice cloning system},
  author={Luong, Hieu-Thi and Yamagishi, Junichi},
  journal={IEEE/ACM Transactions on Audio, Speech, and Language Processing},
  volume={28},
  pages={2967--2981},
  year={2020},
  publisher={IEEE}
}

@misc{tjandra2025metaaudioboxaestheticsunified,
      title={Meta Audiobox Aesthetics: Unified Automatic Quality Assessment for Speech, Music, and Sound}, 
      author={Andros Tjandra and Yi-Chiao Wu and Baishan Guo and John Hoffman and Brian Ellis and Apoorv Vyas and Bowen Shi and Sanyuan Chen and Matt Le and Nick Zacharov and Carleigh Wood and Ann Lee and Wei-Ning Hsu},
      year={2025},
      eprint={2502.05139},
      archivePrefix={arXiv},
      primaryClass={cs.SD},
      url={https://arxiv.org/abs/2502.05139}, 
}

@misc{chen2024socialbenchsocialityevaluationroleplaying,
      title={SocialBench: Sociality Evaluation of Role-Playing Conversational Agents}, 
      author={Hongzhan Chen and Hehong Chen and Ming Yan and Wenshen Xu and Xing Gao and Weizhou Shen and Xiaojun Quan and Chenliang Li and Ji Zhang and Fei Huang and Jingren Zhou},
      year={2024},
      eprint={2403.13679},
      archivePrefix={arXiv},
      primaryClass={cs.CL},
      url={https://arxiv.org/abs/2403.13679}, 
}

@inproceedings{lerner-etal-2022-bazinga,
    title = "Bazinga! A Dataset for Multi-Party Dialogues Structuring",
    author = {Lerner, Paul  and
      Bergo{\"e}nd, Juliette  and
      Guinaudeau, Camille  and
      Bredin, Herv{\'e}  and
      Maurice, Benjamin  and
      Lefevre, Sharleyne  and
      Bouteiller, Martin  and
      Berhe, Aman  and
      Galmant, L{\'e}o  and
      Yin, Ruiqing  and
      Barras, Claude},
    editor = "Calzolari, Nicoletta  and
      B{\'e}chet, Fr{\'e}d{\'e}ric  and
      Blache, Philippe  and
      Choukri, Khalid  and
      Cieri, Christopher  and
      Declerck, Thierry  and
      Goggi, Sara  and
      Isahara, Hitoshi  and
      Maegaard, Bente  and
      Mariani, Joseph  and
      Mazo, H{\'e}l{\`e}ne  and
      Odijk, Jan  and
      Piperidis, Stelios",
    booktitle = "Proceedings of the Thirteenth Language Resources and Evaluation Conference",
    month = jun,
    year = "2022",
    address = "Marseille, France",
    publisher = "European Language Resources Association",
    url = "https://aclanthology.org/2022.lrec-1.367/",
    pages = "3434--3441"
}

@inproceedings{10.1145/3701551.3703583,
author = {Guo, Fang and Li, Wenyu and Zhuang, Honglei and Luo, Yun and Li, Yafu and Yan, Le and Zhu, Qi and Zhang, Yue},
title = {MCRanker: Generating Diverse Criteria On-the-Fly to Improve Pointwise LLM Rankers},
year = {2025},
isbn = {9798400713293},
publisher = {Association for Computing Machinery},
address = {New York, NY, USA},
url = {https://doi.org/10.1145/3701551.3703583},
doi = {10.1145/3701551.3703583},
booktitle = {Proceedings of the Eighteenth ACM International Conference on Web Search and Data Mining},
pages = {944–953},
numpages = {10},
location = {Hannover, Germany},
series = {WSDM '25}
}

@inproceedings{10.1145/3589335.3651584,
author = {Li, Wenyu and Zhu, Yinuo and Lin, Xin and Li, Ming and Jiang, Ziyue and Zeng, Ziqian},
title = {Zero-shot Explainable Mental Health Analysis on Social Media by Incorporating Mental Scales},
year = {2024},
isbn = {9798400701726},
publisher = {Association for Computing Machinery},
address = {New York, NY, USA},
url = {https://doi.org/10.1145/3589335.3651584},
doi = {10.1145/3589335.3651584},
booktitle = {Companion Proceedings of the ACM Web Conference 2024},
pages = {959–962},
numpages = {4},
location = {Singapore, Singapore},
series = {WWW '24}
}

@misc{zhan2025vstylebenchmarkvoicestyle,
      title={VStyle: A Benchmark for Voice Style Adaptation with Spoken Instructions}, 
      author={Jun Zhan and Mingyang Han and Yuxuan Xie and Chen Wang and Dong Zhang and Kexin Huang and Haoxiang Shi and DongXiao Wang and Tengtao Song and Qinyuan Cheng and Shimin Li and Jun Song and Xipeng Qiu and Bo Zheng},
      year={2025},
      eprint={2509.09716},
      archivePrefix={arXiv},
      primaryClass={cs.SD},
      url={https://arxiv.org/abs/2509.09716}, 
}

@misc{hassid2024textuallypretrainedspeechlanguage,
      title={Textually Pretrained Speech Language Models}, 
      author={Michael Hassid and Tal Remez and Tu Anh Nguyen and Itai Gat and Alexis Conneau and Felix Kreuk and Jade Copet and Alexandre Defossez and Gabriel Synnaeve and Emmanuel Dupoux and Roy Schwartz and Yossi Adi},
      year={2024},
      eprint={2305.13009},
      archivePrefix={arXiv},
      primaryClass={cs.CL},
      url={https://arxiv.org/abs/2305.13009}, 
}

@misc{chen2024voicebenchbenchmarkingllmbasedvoice,
      title={VoiceBench: Benchmarking LLM-Based Voice Assistants}, 
      author={Yiming Chen and Xianghu Yue and Chen Zhang and Xiaoxue Gao and Robby T. Tan and Haizhou Li},
      year={2024},
      eprint={2410.17196},
      archivePrefix={arXiv},
      primaryClass={cs.CL},
      url={https://arxiv.org/abs/2410.17196}, 
}

@misc{liu2025vocalbenchbenchmarkingvocalconversational,
      title={VocalBench: Benchmarking the Vocal Conversational Abilities for Speech Interaction Models}, 
      author={Heyang Liu and Yuhao Wang and Ziyang Cheng and Ronghua Wu and Qunshan Gu and Yanfeng Wang and Yu Wang},
      year={2025},
      eprint={2505.15727},
      archivePrefix={arXiv},
      primaryClass={cs.CL},
      url={https://arxiv.org/abs/2505.15727}, 
}

@misc{jiang2025speechrolelargescaledatasetbenchmark,
      title={SpeechRole: A Large-Scale Dataset and Benchmark for Evaluating Speech Role-Playing Agents}, 
      author={Changhao Jiang and Jiajun Sun and Yifei Cao and Jiabao Zhuang and Hui Li and Xiaoran Fan and Ming Zhang and Junjie Ye and Shihan Dou and Zhiheng Xi and Jingqi Tong and Yilong Wu and Baoyu Fan and Zhen Wang and Tao Liang and Zhihui Fei and Mingyang Wan and Guojun Ma and Tao Ji and Tao Gui and Qi Zhang and Xuanjing Huang},
      year={2025},
      eprint={2508.02013},
      archivePrefix={arXiv},
      primaryClass={cs.CL},
      url={https://arxiv.org/abs/2508.02013}, 
}

\section{Appendix}
\label{sec:appendix}

\subsection{Prompts for Generation and Evaluation}
\label{app:prompts}

To ensure reproducibility, we provide the exact prompts used for baseline generation and automated evaluation metrics.

\begin{table}[h]
    \centering

    \begin{tabularx}{\linewidth}{X}
        \toprule
        \textbf{Prompt for Baseline Models (like GPT-4o-Audio)} \\
        \midrule
        Please imitate the character of ``\texttt{\{role\}}'' in TV series of ``\texttt{\{TV\}}'' and reply to the content of this audio given to you in ``\texttt{\{role\}}'' 's style. 
        
         \\
        
        Directly generate your answer, don't say any other words. \\
        \bottomrule
    \end{tabularx}
    \caption{Generation prompt of baseline models.}
    \label{tab:prompt_baseline}
\end{table}

\begin{table}[h]
    \centering

    \begin{tabularx}{\linewidth}{X}
        \toprule
        \textbf{Prompt for Content Quality (CQ) Evaluation} \\
        \midrule
        I will give you one turn dialogue, the first sentence is said by some one else and the second sentence is said by ``\texttt{\{role\}}'' from the TV series ``\texttt{\{TV\}}''. After carefully read, you should score the helpfulness of the second sentence to evaluate under the character of ``\texttt{\{role\}}'' whether it is a good answer to the first sentence. 
        
        \\
        
        The first sentence is: ``\texttt{\{gold\_transcription\}}''
        
        \\
        
        The second sentence from ``\texttt{\{role\}}'' is: ``\texttt{\{transcription\}}''
        
        \\
        
        Your helpfulness score should be from 0 (lowest) to 2 (highest). 
                
         \\
        
        Directly generate your score in an int number, don't say any other words. \\
        \bottomrule
    \end{tabularx}
    \caption{Evaluation prompt of CQ.}
    \label{tab:prompt_cq}
\end{table}

\begin{table}[h]
    \centering

    \begin{tabularx}{\linewidth}{X}
        \toprule
        \textbf{Prompt for Content Personalization (CP) Evaluation} \\
        \midrule
        I will give you two audios, one of them is the audio from ```\texttt{\{role\}}'' in the TV series ``\texttt{\{TV\}}'' and the other audio is from UNKNOWN. 
                
         \\
        
        Please help me determine whether the two audios maintain the same speaking style and whether their content is obviously spoken by the same character ``\texttt{\{role\}}''. Please score the UNKNOWN audio's Content Personalization degree from 0 (lowest) to 2 (highest). 
                
         \\
        
        Directly generate your score in an int number, don't say any other words. \\
        \bottomrule
    \end{tabularx}
    \caption{Evaluation prompt of CP.}
    \label{tab:prompt_cp}
\end{table}

\end{document}